\documentclass[aps,prb,twocolumn,a4]{revtex4-1} 
\pdfoutput=1
\usepackage{amsmath,amssymb,amsfonts,upgreek}
\usepackage{color}
\usepackage{graphicx}
\usepackage{setspace}
\usepackage{bm}
\usepackage{comment}
\usepackage{hyperref}

\graphicspath{{./Figures/}}

\begin{document}

\title{Tuning the metal-insulator transition in $d^1$ and $d^2$
  perovskites by epitaxial strain: a first principles-based study}

\author{Gabriele Sclauzero}
\affiliation{Materials Theory, ETH Z\"urich, Wolfgang-Pauli-Strasse
  27, 8093 Z\"urich, Switzerland}
\author{Krzysztof Dymkowski}
\affiliation{Materials Theory, ETH Z\"urich, Wolfgang-Pauli-Strasse
  27, 8093 Z\"urich, Switzerland}
\author{Claude Ederer}
\email{claude.ederer@mat.ethz.ch}
\affiliation{Materials Theory, ETH Z\"urich, Wolfgang-Pauli-Strasse
  27, 8093 Z\"urich, Switzerland}

\date{\today}

\begin{abstract}
We investigate the effect of epitaxial strain on the Mott
metal-insulator transition (MIT) in perovskite systems with $d^1$ and
$d^2$ electron configuration of the transition metal (TM) cation. We
first discuss the general trends expected from the changes in the
crystal-field splitting and in the hopping parameters that are induced
by epitaxial strain. We argue that the strain-induced crystal-field
splitting generally favors the Mott-insulating state, whereas the
strain-induced changes in the hopping parameters favor the metallic
state under compressive strain and the insulating state under tensile
strain.  Thus, the two effects can effectively cancel each other under
compressive strain, while they usually cooperate under tensile strain,
in this case favoring the insulating state.
We then validate these general considerations by performing electronic
structure calculations for several $d^1$ and $d^2$ perovskites, using
a combination of density functional theory (DFT) and dynamical
mean-field theory (DMFT). We isolate the individual effects of
strain-induced changes in either hopping or crystal-field by
performing DMFT calculations where we fix one type of parameter to the
corresponding unstrained DFT values. These calculations confirm our
general considerations for SrVO$_3$ ($d^1$) and LaVO$_3$ ($d^2$),
whereas the case of LaTiO$_3$ ($d^1$) is distinctly different, due to
the strong effect of the octahedral tilt distortion in the underlying
perovskite crystal structure.
Our results demonstrate the possibility to tune the electronic
properties of correlated TM oxides by using epitaxial strain, which
allows to control the strength of electronic correlations and the
vicinity to the Mott MIT.
\end{abstract}

\maketitle

\section{Introduction}

Using modern growth techniques, it is nowadays possible to create
high-quality epitaxial thin films and heterostructures of complex
transition metal (TM) oxides with well-defined composition and
atomically sharp interfaces. Thereby, epitaxial strain, which is
caused by the lattice mismatch between thin film and substrate
materials, has emerged as a very effective tool to design and optimize
specific functional
properties.~\cite{Schlom_et_al:2008,Li/Shan/Ma:2014,Schlom_et_al:2014}
For example, epitaxial strain has been shown to strongly affect
ferroelectric Curie temperatures and
polarization,\cite{Pertsev/Zembilgotov/Tagantsev:1998,Choi_et_al:2004,Ederer/Spaldin_PRL:2005}
and even induce ferroelectricity in otherwise non-ferroelectric
materials.\cite{Haeni_et_al:2004,Lee_et_al:2010} In magnetic
materials, epitaxial strain can be used to tune magnetic anisotropy
and switch between in-plane and out-of-plane
anisotropy.\cite{Suzuki_et_al:1999,Heuver_et_al:2015} Furthermore,
strain also affects ionic transport and catalytic
properties~\cite{Yildiz:2014} and allows to tune electronic band
gaps~\cite{Berger/Fennie/Neaton:2011,Choi/Lee:2015} or Fermi
surfaces.\cite{Yoo_et_al:2015,Burganov_et_al:2016}

Particularly interesting are also systems where epitaxial strain can
induce a metal-insulator transition
(MIT).~\cite{Imada/Fujimori/Tokura:1998} Examples of systems where
strain-induced MITs have been reported include nickelates,
\cite{Liu_et_al:2010,Catalano_et_al:2014}
iridates,~\cite{Gruenewald_et_al:2014} and
titanates.~\cite{He_et_al:2012,Dymkowski/Ederer:2014,Yoshimatsu_et_al:2016}
Incorporation of such materials in thin films and heterostructures
allows to tune the characteristics of the MIT and offers the
perspective for novel electronic devices based on the resulting
orders-of-magnitude changes in electrical, thermal, and optical
properties.\cite{Takagi/Hwang:2010,Yang/Ko/Ramanathan:2011}

We have recently shown, using first principles electronic structure
calculations combined with dynamical mean-field theory
(DMFT),\cite{Georges_et_al:1996,Held:2007} that LaTiO$_3$, which in
its unstrained bulk form is a Mott
insulator,\cite{Fujimori_et_al:1992,Arima/Tokura/Torrance:1993,Imada/Fujimori/Tokura:1998}
becomes metallic under a compressive strain of around
$-2$\,\%.\cite{Dymkowski/Ederer:2014} On the other hand, similar
calculations indicate that the Mott-insulating character of the
closely-related material LaVO$_3$ is much less affected by strain, and
that LaVO$_3$ remains insulating under both compressive and tensile
strain.\cite{Sclauzero/Ederer:2015}

These computational results are in good agreement with experimental
studies that observe metallic, bulk-like, conductivity in
compressively strained thin films of LaTiO$_3$ grown on
SrTiO$_3$.\cite{Wong_et_al:2010,He_et_al:2012} Similar experiments for
thin films of LaVO$_3$ grown on SrTiO$_3$ also observe metallic
conductivity, however, in this case the conductivity seems to be
restricted to the interface region between the thin film and the
substrate, which indicates that the metallic character of LaVO$_3$
thin films is not due to epitaxial strain.\cite{He_et_al:2012}

It is important to note that many different factors can play a role in
determining the properties of oxide thin films and heterostructures.
In particular, metallic behavior of otherwise insulating materials can
be caused by several different effects. Apart from epitaxial strain,
important factors are structural and electronic reconstruction at the
interface,\cite{Okamoto/Millis:2004,Nakagawa/Hwang/Mueller:2006} the
specific interface
chemistry,\cite{Nakagawa/Hwang/Mueller:2006,Willmott_et_al:2007}
confinement effects,\cite{Yoshimatsu_et_al:2010,Gu_et_al:2013} or
defects.\cite{Herranz_et_al:2007,Basletic_et_al:2008,Aschauer_et_al:2013}
The interplay between these effects as well as their relative
importance is generally not known {\it a priori}. However, our
previous work on LaTiO$_3$ has clearly shown that epitaxial strain can
be a major factor that needs to be taken into account to correctly
interpret experimental observations.

In view of this, and considering the different strain responses of
closely related materials such as LaTiO$_3$ and LaVO$_3$, it is
desirable to build up a comprehensive understanding of how epitaxial
strain affects the electronic properties in early TM perovskites, and
in particular their tendency to form a Mott-insulating state. To this
end, here, we examine the effect of epitaxial strain on the MIT in
perovskite TM oxides with $d^1$ and $d^2$ electron configurations on
the TM cation, such as LaTiO$_3$, SrVO$_3$, and LaVO$_3$. As outlined
below, all these systems can be described by an effective
three-orbital model with different integer occupations and are
therefore well suited for a systematic study.

\begin{figure}
\includegraphics[width=0.45\columnwidth]{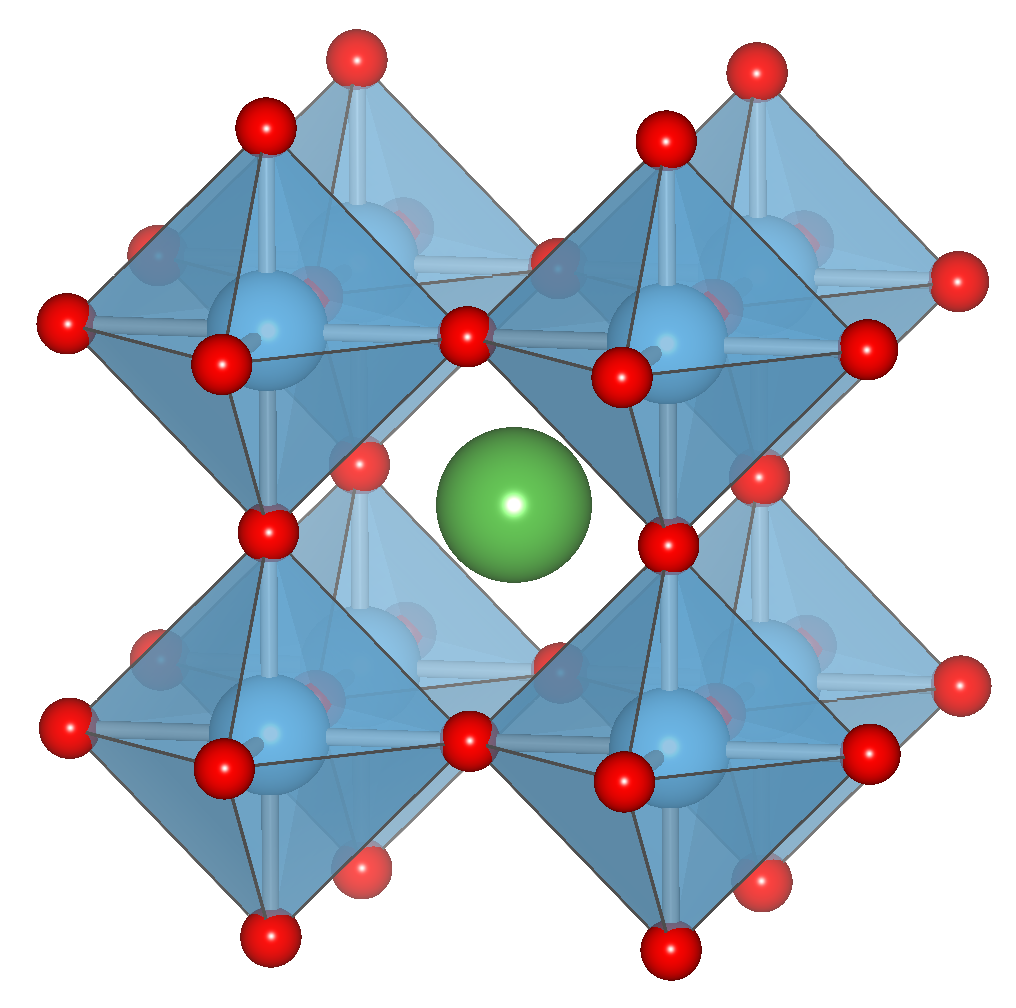}
\quad
\includegraphics[width=0.45\columnwidth]{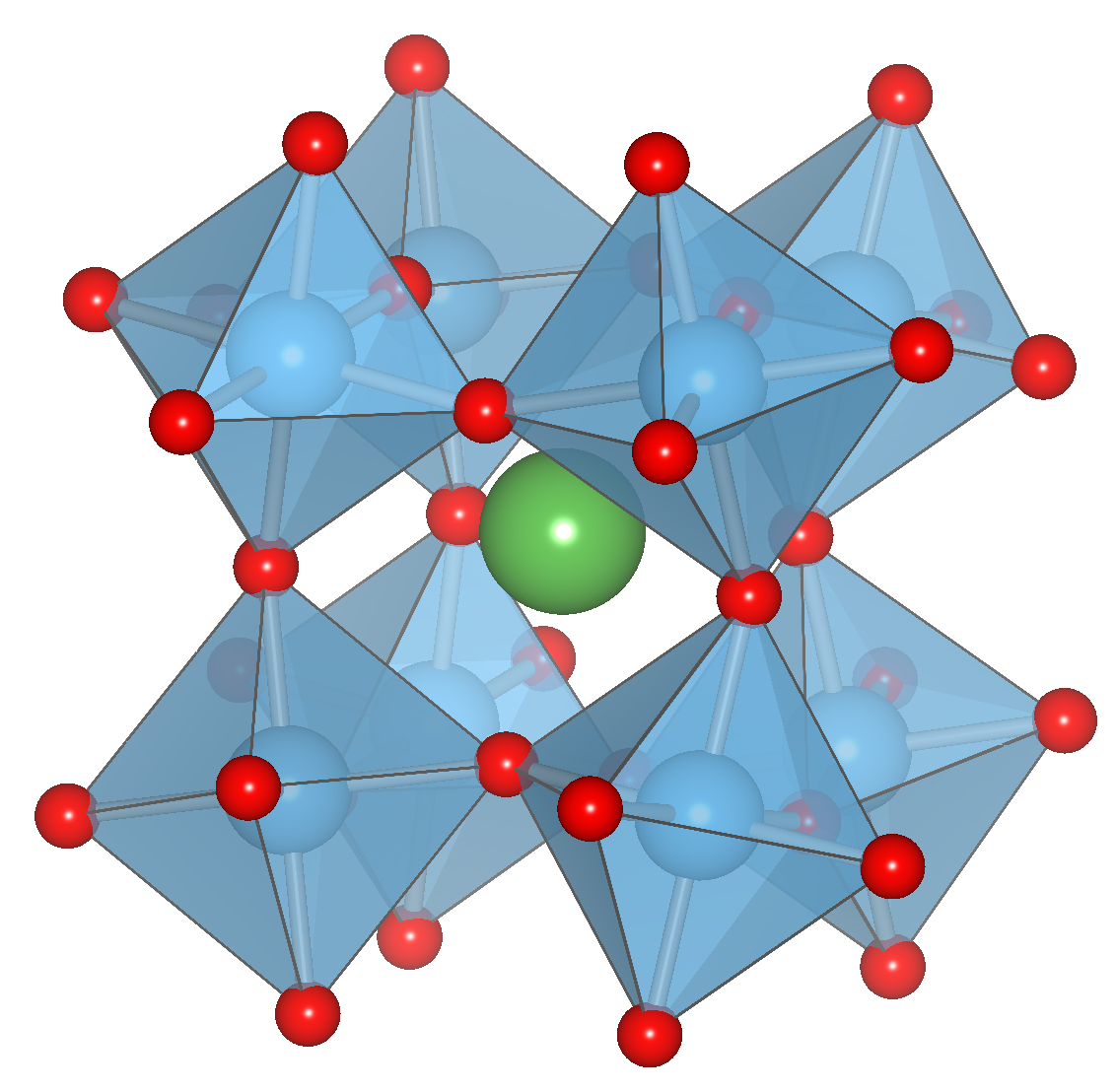}
\caption{Left: Ideal cubic perovskite structure. The TM cations (blue)
  form a simple cubic lattice and are octahedrally coordinated by
  anions (red). The larger $A$-site cations (green) occupy the voids
  formed in between eight anion octahedra. Right: Distorted perovskite
  structure exhibiting tilted octahedra. The depicted case corresponds
  to $Pbnm$ space group symmetry.}
\label{fig:Pbnm}
\end{figure}

Within the perovskite crystal structure, the TM cation is octahedrally
coordinated by oxygen ligands (see Fig.~\ref{fig:Pbnm}). Thus, the
$d$-orbital manifold of the TM cation is split by the octahedral
component of the crystal-field into three $t_{2g}$ and two $e_g$
orbitals.\footnote{In most cases the actual site symmetry is lower
  than cubic, due to the presence of small structural distortions, in
  particular rotations of the oxygen octahedra around the central TM
  cation. In spite of such symmetry-lowering, in the following we are
  using the labels ``$t_{2g}$'' and ``$e_g$'' to denote subsets of $d$
  orbitals.}  Hybridization of the TM $t_{2g}$ orbitals with the $p$
orbitals of the surrounding oxygen ligands leads to the formation of a
partially-filled group of bands, in the following denoted ``$t_{2g}$
bands'', which governs the low energy electronic properties of the
early TM perovskites with one or two $d$ electrons on the TM
cation.~\cite{Pavarini_et_al:2004,Dang_et_al:2014}

In the following, we first discuss some simple ideas on how strain is
expected to affect the electronic structure of these $t_{2g}$ bands,
and what general trends can be expected from this. After this general
discussion, we present a detailed study of the effect of strain in the
prototypical system SrVO$_3$, which in the bulk exhibits a perfect
cubic perovskite structure without octahedral rotations (see
Fig.~\ref{fig:Pbnm}). We analyze the effect of strain on the $t_{2g}$
bands of SrVO$_3$ in terms of crystal-field splittings and hopping
amplitudes. These quantities are obtained by constructing maximally
localized Wannier functions corresponding to the $t_{2g}$ bands,
obtained for the strained structures within Kohn-Sham density
functional theory (DFT). Then, we perform DMFT calculations based on
these electronic bands and monitor the effect of strain on the
critical interaction strength for the Mott MIT.

In order to clearly distinguish the effect of the strain-induced
crystal-field splittings between the $t_{2g}$ orbitals from the effect
of the strain-induced changes in the hopping amplitudes, and to make a
better connection to our previous results for LaTiO$_3$ and LaVO$_3$,
we perform additional DMFT calculations for all three materials, where
we fix either the hopping amplitudes or the crystal-field splitting to
the corresponding unstrained values and only consider the
strain-induced changes in the respective other quantity. Finally, this
allows us to formulate some general trends, applicable to other
perovskites with $d^1$ and $d^2$ electron configuration of the TM
cation, which can provide useful guidance for future studies of these
materials in thin films and as components of oxide hetrostructures

\section{General considerations on the effect of epitaxial strain}
\label{sec:general}

If a material is epitaxially grown on a substrate without forming
dislocations, then the extensions of the crystallographic unit cell in
the plane parallel to the substrate surface are constrained to the
corresponding lattice constants of the substrate, whereas the unit
cell can freely adjust in the perpendicular direction. This elastic
deformation of the unit cell will affect bond distances and,
potentially, bond angles even far away from the substrate-film
interface.

In this section, we discuss the expected effect of these structural
changes on the electronic properties of early transition metal
perovskites, using a tight-binding (TB) description for the
partially-filled $t_{2g}$ bands formulated in a basis of ``effective''
$t_{2g}$ orbitals centered on the TM cations. Such a TB description
can be obtained, e.g., by constructing maximally localized Wannier
functions from the corresponding Kohn-Sham
bands.\cite{Marzari_et_al:2012,Lechermann_et_al:2006} The resulting
$t_{2g}$-like Wannier functions are typically more extended than
atomic orbitals and also contain contributions on the surrounding
oxygen sites stemming from hybridization between atomic-like cation
$d$ orbitals with oxygen $p$ orbitals (see, e.g.,
Refs.~\onlinecite{Lechermann_et_al:2006,Dang_et_al:2014,Dymkowski/Ederer:2014}).

The general form of such a TB Hamiltonian is as follows:
\begin{align}
H_0 & = \sum_{I,m,n} \varepsilon^{I}_{nm} d_{In}^\dagger d_{Im}
\nonumber \\* & + \sum_{I,J,m,n} t^{IJ}_{nm} ( d^\dagger_{In} d_{Jm} +
d^\dagger_{Jm}d_{In}) \quad .
\label{eq:TB}
\end{align}
Here, $I$ and $J$ indicate different TM sites, $m$ and $n$ indicate
different orbitals centered at these sites, and $d^\dagger_{In}$ is
the creation operator for an electron on site $I$ in orbital
$n$. Since there is no explicit spin dependence in Eq.~\eqref{eq:TB},
the spin index has been suppressed for more clarity.

The first term in Eq.~\eqref{eq:TB} contains the on-site
\emph{crystal-field energies}, $\varepsilon^{I}_{nm}$, which can be
obtained as matrix elements of the Hamiltonian between Wannier
orbitals centered on the same site $I$. The second term in
Eq.~\eqref{eq:TB} contains the inter-site \emph{hopping amplitudes},
$t^{IJ}_{nm}$, which are obtained as matrix elements of the
Hamiltonian between Wannier orbitals centered on different sites $I$
and $J$. In general, the strain-induced changes in bond lengths and
bond angles affect both the crystal-field energies and the hopping
parameters.

In order to describe a Mott insulator (or a metallic system close to a
Mott-insulating state), the non-interacting Hamiltonian in
Eq.~\eqref{eq:TB} has to be supplemented by a term representing the
on-site electron-electron interaction. We use the so-called
Slater-Kanamori form (see, e.g.,
Ref.~\onlinecite{Werner/Gull/Millis:2009}):
\begin{align}
&H_\text{int}=\sum_{n} U n_{n,\uparrow} n_{n,\downarrow}+\sum_{n\ne
    m,\sigma} U' n_{n,\sigma} n_{m,-\sigma} \nonumber\\ &+\sum_{n\ne
    m,\sigma} (U'-J) n_{n,\sigma}n_{m,\sigma}\nonumber\\ &-\sum_{n\ne
    m}J(d^\dagger_{n,\downarrow}d^\dagger_{m,\uparrow}d_{m,\downarrow}d
  _{n,\uparrow} +
  d^\dagger_{m,\uparrow}d^\dagger_{m,\downarrow}d_{n,\uparrow}d_{n,\downarrow}
  + h.c.) \ .
\label{eq:slater-kanamori}
\end{align}
Here, $d^\dagger_{n,\sigma}$ is the creation operator for an electron
in orbital $n$ with spin $\sigma$, and
$n_{n,\sigma}=d^\dagger_{n,\sigma}d_{n,\sigma}$. The parameters $U$,
$U'$, and $J$ describe the strength of the intra- and inter-orbital
electron-electron interaction and the Hund's rule coupling,
respectively, and $U'=U-2J$. The site index $I$ has been suppressed in
Eq.~\eqref{eq:slater-kanamori}, since all terms are purely local.

\subsection{Crystal-field splitting}
\label{subsec:CF}

We first discuss the effect of the strain-induced crystal-field
splitting between the effective $t_{2g}$ Wannier orbitals. We consider
the case of an ideal perovskite structure without octahedral rotations
(see Fig.~\ref{fig:Pbnm}(a)), and we assume that the surface of a
hypothetical substrate can be represented by a two-dimensional square
lattice oriented parallel to the (001) plane ($x$-$y$ plane) of the
perovskite. We further assume that the thin film adopts the in-plane
lattice constant of the underlying substrate whereas the out-of-plane
lattice constant adjusts to minimize the elastic energy of the
system. In general, this will lead to a tetragonal deformation of the
perovskite unit cell.

\begin{figure}
\includegraphics[width=0.75\columnwidth]{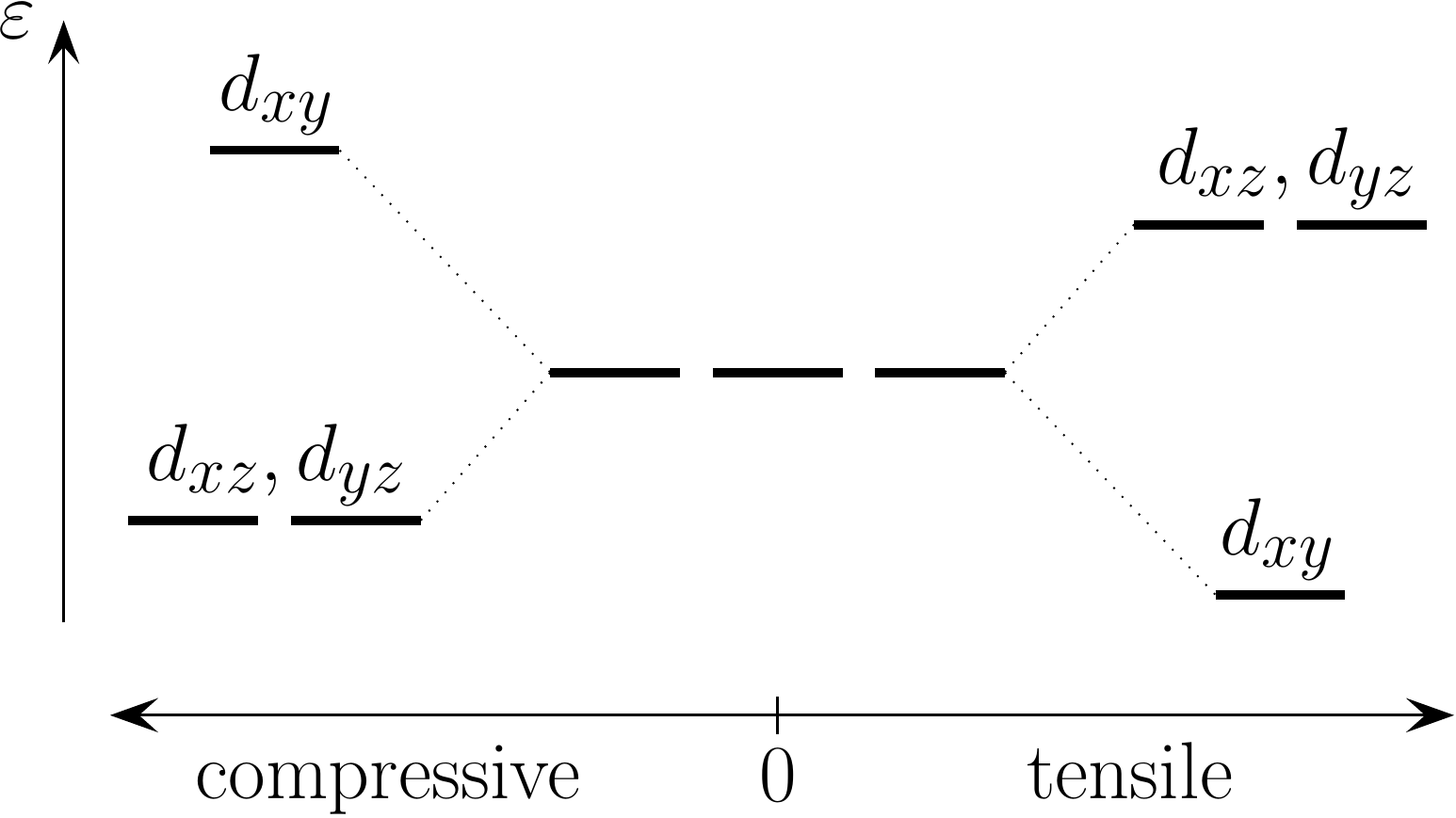}
\caption{Schematic depiction of the crystal-field splitting between
  $t_{2g}$ orbitals induced under compressive and tensile epitaxial
  strain, respectively. Here, the $x$-$y$ plane is oriented parallel
  to the surface of the hypothetical substrate.}
\label{fig:crystal-field-schematic}
\end{figure}

The resulting crystal-field energies in Eq.~\eqref{eq:TB} are diagonal
and do not depend on the site index $I$, i.e. $\varepsilon^{I}_{nm} =
\varepsilon_m \delta_{nm}$.  The crystal-field associated to the
epitaxial strain reduces the symmetry of the perovskite crystal
structure from cubic to tetragonal, lifting the three-fold degeneracy
among the $t_{2g}$ orbitals into a doubly-degenerate ($d_{xz}$ and
$d_{yz}$) and a non-degenerate ($d_{xy}$) set of orbitals, as shown
schematically in Fig.~\ref{fig:crystal-field-schematic}.
Under compressive strain, the oxygen ligands within the $x$-$y$ plane
move closer to the central TM cation. This strongly increases the
hybridization of the $d_{xy}$ orbital with its surrounding ligand
O-$p$ states, thus shifting the $d_{xy}$ orbital to higher energies
relative to the $d_{xz}/d_{yz}$ orbitals due to the antibonding
character of the $t_{2g}$ states. Analogously, under tensile strain,
the increase of the TM-O bond distances within the $x$-$y$ plane
lowers the energy of the $d_{xy}$ orbital relative to $d_{xz}/d_{yz}$.

The effect of such a crystal-field splitting on the Mott MIT in a
simple three band Hubbard model, applicable to systems with partially
filled $t_{2g}$ states, has been studied in
Ref.~\onlinecite{Werner/Gull/Millis:2009} (see also
Refs.~\onlinecite{Kita/Ohashi/Kawakami:2011,Huang/Du/Dai:2012}). It
was found that for both one- and two-electron-filling, i.e.,
corresponding to a $d^1$ and $d^2$ electron configuration of the TM
cation, the crystal-field splitting reduces the critical value for the
Hubbard $U$ parameter that is required to obtain an insulating
state. This can be understood from the fact that the splitting reduces
the orbital degeneracy, and, as shown in
Refs.~\onlinecite{Gunnarsson/Koch/Martin:1996,Florens_et_al:2002}, the
critical $U$ decreases with decreasing orbital degeneracy. A
particularly strong reduction of the critical $U$ was found for the
case with two orbitals at lower energy and one orbital at higher
energy, i.e., corresponding to compressive strain, and two electrons
per site.~\cite{Werner/Gull/Millis:2009} In this case, the two
lower-lying orbitals are effectively half-filled,
and the Hund's rule interaction between the two electrons,
represented by the parameter $J$ in \eqref{eq:slater-kanamori},
leads to a strong stabilization of the Mott insulating state.
%
Thus, a strain-induced crystal-field splitting is generally expected
to favor the insulating state, with a particularly strong effect for a
$d^2$ system under compressive strain. 

\subsection{Hopping amplitudes}
\label{subsec:hopping}

While the effect of a crystal-field on the Mott transition has already
been studied within a simplified three orbital Hubbard
model,\cite{Werner/Gull/Millis:2009,Kita/Ohashi/Kawakami:2011,Huang/Du/Dai:2012}
we are not aware of any systematic studies on how the strain-induced
modifications of the hopping parameters, $t^{IJ}_{nm}$ in
Eq.~\ref{eq:TB}, will affect the MIT. In the following, we therefore
discuss how the hopping parameters are affected by epitaxial strain,
and what resulting effects on the electronic properties can be
expected.

We first note that the hopping between two effective $t_{2g}$ Wannier
functions located at adjacent TM sites should be viewed as an
\emph{effective} hopping process that is mediated by the oxygen anion
situated between the two TM cations, i.e., an electron first hops from one
of the TM sites into a $p$ orbital on the oxygen site and then onto
the other TM site. Thus, the effective $t_{2g}$-$t_{2g}$ hopping
amplitude is determined mostly by the TM-O bond length and by the
TM-O-TM bond angle, whereas the direct TM-TM distance is less
important.

Compressive epitaxial strain reduces the TM-O bond distances in the
two in-plane directions, i.e., parallel to the film-substrate
interface, and increases the bond distance in the perpendicular
direction, due to the outward relaxation of the out-of-plane lattice
parameter. This leads to increased in-plane hopping amplitudes and
reduced out-of-plane hoppings compared to the unstrained case. Tensile
epitaxial strain has the opposite effect.

On the other hand, if octahedral rotations are present in the
structure (see Fig.~\ref{fig:Pbnm}), as is the case for most
perovskites, then the TM-O-TM bond angles are distorted from the ideal
value of 180$^\circ$ and the amount of this distortion will change
with strain. In general, the in-plane bond angles will become more
distorted under compressive strain, whereas the out-of plane bonds
will be straightened out,\footnote{We note that in the presence of
  octahedral rotations the TM-O bonds are not exactly parallel or
  perpendicular to the substrate plane. Nevertheless, for typical
  octahedral rotation angles smaller than 15-20$^\circ$, a clear
  distinction between in-plane and out-of-plane oriented bonds is
  easily possible.}  i.e. the corresponding bond angles will become
less distorted (see, e.g.,
Refs.~\onlinecite{Rondinelli/Spaldin:2011,Dymkowski/Ederer:2014,Sclauzero/Ederer:2015}). Since
the hopping amplitudes decrease with increasing distortion of the bond
angles, i.e., with increasing deviation from the ideal value of
180$^\circ$, this leads to decreasing in-plane hopping amplitudes and
increasing out-of-plane hoppings under compressive strain. Again,
tensile strain leads to the opposite trends.

Thus, it can be seen that the trends expected from the strain-induced
changes in the TM-O-TM bond angles are exactly opposite to those
expected from the strain-induced changes in the bond distances. Our
previous calculations for LaTiO$_3$ and LaVO$_3$ indicate that in both
materials the influence of the bond distances dominates, i.e., the
in-plane hoppings are increased under compressive strain and the
out-of-plane hoppings are decreased (and vice versa for tensile
epitaxial strain).\cite{Dymkowski/Ederer:2014,Sclauzero/Ederer:2015}

How will these changes of the hopping amplitudes affect the Mott MIT?
For the simplest case with one orbital per site, the Mott transition
is governed by the ratio $U/W$, where $W$ is the corresponding
bandwidth.\cite{Georges_et_al:1996} The present case with three
orbitals is more complex. However, to a good approximation, the
$t_{2g}$ bands can be viewed as three independent bands corresponding
to $d_{xy}$, $d_{xz}$, and $d_{yz}$ orbitals, respectively. From the
above considerations it therefore follows that the bandwidth of the
$d_{xy}$-derived band, which is determined by the in-plane hopping,
will increase under compressive strain and decrease under tensile
strain. On the other hand, the $d_{xz}$- as well as the
$d_{yz}$-derived bands will become more anisotropic but, to a first
approximation, the total width of these bands will only be weakly
affected by epitaxial strain, since the increase of the in-plane
hopping under compressive strain will, at least partially, be
compensated by the decrease in the out-of-plane hopping.
Overall, these strain-induced modifications of the nearest neighbor
hopping amplitudes lead to a moderate increase of the total $t_{2g}$
bandwidth under compressive strain and to a slight decrease of
bandwidth under tensile strain, as also confirmed by our previous DFT
calculations for LaTiO$_3$ and
LaVO$_3$.\cite{Dymkowski/Ederer:2014,Sclauzero/Ederer:2015}

Thus, one can expect that the increase in bandwidth under compressive
strain will favor the metallic state, whereas the reduced bandwidth
under tensile strain will be more favorable for the Mott-insulating
state. 
However, it is not clear {\it a priori} how the strain-induced
anisotropy in the $d_{xz}$- and $d_{yz}$-derived bands will affect the
MIT. Systematic model calculations are required to explore this issue.
Furthermore, the difference in bandwidth for the $d_{xy}$-derived band
compared to the other two bands could potentially lead to
orbitally-selective Mott
transitions.~\cite{Anisimov_et_al:2002,Koga_et_al:2004} Another
effect, which, however, is not straightforward to incorporate
systematically into a simplified TB model, is that the presence of
octahedral tilts leads to mixing between the three bands corresponding
to the three different orbital characters. Again, it is unclear how
such intermixing will affect the MIT.

In this work we are not aiming for a full clarification of all these
issues. Instead, we verify the simple general considerations outlined
in this section using realistic first principles-based electronic
structure calculations for different $d^1$ and $d^2$ perovskite TM
oxides under epitaxial strain. Before presenting our results, we
briefly summarize the (expected) net effect of the strain-induced
changes in crystal-field splitting and hopping amplitudes on the MIT.

\subsection{Expected strain dependence of the MIT}
\label{subsec:summary}

As outlined in the preceeding subsections, the strain-induced
crystal-feld splitting will always lower the critical $U$ for the Mott
transition and promote the insulating phase. On the other hand, we
expect the strain-related changes in the hopping amplitudes to
increase the critical $U$ under compressive strain and decrease it
under tensile strain. Thus, partial cancelation between the
crystal-field- and hopping-related effects can occur under compressive
strain, whereas tensile strain is expected to promote insulating
behavior in all cases.

The opposing trends resulting from crystal-field splitting and hopping
amplitudes under compressive strain, have already been suggested as
explanation for the weak effect of compressive strain on the MIT in
the $d^2$ system LaVO$_3$.~\cite{Sclauzero/Ederer:2015}.
On the other hand, it has been found that tensile strain strongly
reinforces the insulating character of the $d^1$ Mott insulator
LaTiO$_3$,~\cite{Dymkowski/Ederer:2014} and also increases the
critical $U$ in the $d^2$ system
LaVO$_3$.~\cite{Sclauzero/Ederer:2015}
Furthermore, the discussed trends also indicate that a metallic $d^1$
or $d^2$ system, such as, e.g., SrVO$_3$, is expected to move closer
to the Mott-insulating state, and might even become insulating under
strong tensile strain.

We note that in our discussion we have assumed that the interaction
parameters, $U$ and $J$, are not affected by the epitaxial
strain. Thus, we assume that the screening of the electron-electron
interaction is not significantly affected by the strain-induced
structural modifications. Even though this might indeed be a good
approximation, the corresponding quantitative changes remain to be
verified.

\section{Computational method}

To validate the general considerations outlined in the previous
section, we perform electronic structure calculations for a set of
representative materials using density functional theory
(DFT)~\cite{Hohenberg/Kohn:1964,Kohn/Sham:1965} in combination with
dynamical mean-field theory (DMFT).~\cite{Georges_et_al:1996}

We address the effect of epitaxial strain, by using bulk unit cells
with periodic boundary conditions in all three dimensions, where we
constrain the lattice parameters in the two directions corresponding
to the substrate plane, while relaxing all other structural degrees of
freedom. Thus, we do not explicitly consider a substrate in our
calculations, e.g., by using a slab geometry and large
supercells. Consequently, our approach allows to clearly distinguish
the bulk-like strain effect from other factors related to the
interface between the thin film material and the substrate.

Most systems investigated in this work exhibit a distorted perovskite
structure with space group symmetry $Pbnm$ in their bulk forms. We
consider the case where the substrate presents a square lattice on its
surface, and we assume a growth geometry where the two shorter lattice
vectors of the orthorhombic $Pbnm$ structure are parallel to the
surface plane of the substrate and are constrained to have equal
length. The longest lattice vector of the $Pbnm$ structure is then
oriented perpendicular to the substrate plane and is allowed to adjust
its length in order to minimize the energy of the system under the
epitaxial constraint. Simultaneously, all internal structural degrees
of freedom related to the individual atomic positions are also
relaxed. The chosen geometry preserves the $Pbnm$ symmetry of the bulk
system and allows for a systematic comparison between the different
materials considered in this work. 
The case of SrVO$_3$, which in its bulk form exhibits a perfect cubic
perovskite structure with $Pm\bar{3}m$ space group symmetry, is
treated analogously, i.e. assuming growth along the [001] direction on
a square lattice substrate.

The applied strain is defined as $s = (a-a_0)/a_{0}$, where $a$ is the
constrained in-plane lattice constant (corresponding to the surface
lattice constant of the hypothetical substrate) and $a_{0}$ is the
unstrained reference lattice constant of the thin film material. For
SrVO$_3$, $a_{0}$ is the theoretical equilibrium lattice constant of
the ideal perovskite ($Pm\bar{3}m$) structure, while for the
orthorhombic $Pbnm$ systems LaTiO$_3$ and LaVO$_3$, $a_0$ was taken as
the in-plane lattice parameter that minimizes the total-energy under
the epitaxial constraint (i.e., with
$a=b$).~\cite{Dymkowski/Ederer:2014}

DFT calculations within the generalized gradient approximation
according to Perdew, Burke, and Ernzerhof (PBE)
\cite{Perdew/Burke/Ernzerhof:1996} allow us to relax the crystal
structure under the epitaxial constraint and to obtain the
corresponding electronic band-structure.  We employ the
Quantum\-ESPRESSO (QE) package~\cite{Giannozzi_et_al:2009} with
ultrasoft pseudopotentials~\cite{Vanderbilt:1990} from the QE website.
Semicore states of the different cations are included in the valence
($3s$ and $3p$ for V and Ti, $4s$ and $4p$ for Sr, $5s$ and $5p$ for
La), while projectors for the empty La-$4f$ shell are not included in
the potential. The plane wave cutoffs used to represent the wave
functions (charge density) are: 60~Ry (500~Ry) for SrVO$_3$, 40~Ry
(480~Ry) for LaTiO$_3$, and 40~Ry (300~Ry) for LaVO$_3$. The Brillouin
zone was sampled with a regular $k$-point grid with dimensions $6
\times 6 \times 4$ for LaTiO$_3$ and LaVO$_3$ and $8 \times 8 \times
8$ for SrVO$_3$.

We note that, since we are interested in the paramagnetic structures
at room temperature, we are performing non-spin-polarized
calculations. In general, we obtain good agreement with the known bulk
structures for all materials studied in this work (see also
Refs.~\onlinecite{Dymkowski/Ederer:2014,Sclauzero/Ederer:2015}). We
note that these crystal structures do not exhibit any distortions that
are specifically driven by the electron-electron interaction (e.g.,
Jahn-Teller distortions) and thus a PBE treatment is sufficient to
obtain accurate structural properties.

After relaxing the structure under the epitaxial constraint,
corresponding to different in-plane lattice constants, we construct a
representation of the electronic bands with predominant TM-$t_{2g}$
orbital character using maximally localized Wannier functions
(MLWFs).~\cite{Marzari_et_al:2012,Mostofi_et_al:2008} 
The MLWFs are constructed from initial projections on $d_{xy}$,
$d_{xz}$, and $d_{yz}$-type orbitals, and we always use a coordinate
system where the cartesian axis are approximately oriented along the
direction of the TM-O bonds, with the $z$ axis perpendicular to the
surface of the hypothetical substrate.

The Kohn-Sham Hamiltonian for the $t_{2g}$-bands, expressed in the
basis of MLWFs, has precisely the form of Eq.~\eqref{eq:TB}, and is
used as noninteracting part of a multi-band Hubbard Hamiltonian, where
the Slater-Kanamori form, Eq.~\eqref{eq:slater-kanamori}, is used to
describe the Coulomb interaction between electrons on the same site.
We then perform DMFT calculations for this
Hamiltonian.~\cite{Georges_et_al:1996} The effective impurity problem
obtained within DMFT is solved using a continuous time hybridization
expansion quantum Monte Carlo solver~\cite{Gull_et_al:2011}
implemented within the TRIQS
library.\cite{Parcollet_et_al:2015,Seth_et_al:2016,Aichhorn_et_al:2016}
All DMFT calculations are performed for an inverse temperature of
$\beta=1/(k_\text{B}T) = 40\, \mathrm{eV}^{-1}$, corresponding to
approximately room temperature. The parameter $U$ in
Eq.~\eqref{eq:slater-kanamori} is varied in order to identify the
critical value for the MIT for each strain, whereas $J$ is fixed to
0.65\,eV, which is a typical value for the materials studied
here.~\cite{Pavarini_et_al:2004,Dymkowski/Ederer:2014,Dang_et_al:2014,Sclauzero/Ederer:2015}
Orbital off-diagonal elements of the impurity self-energy are included
in the calculations. For more details, we refer to the supplemental
material of Ref.~\onlinecite{Dymkowski/Ederer:2014}, where an
analogous setup has been used. From the DMFT calculations we obtain
the local imaginary time Green's function $G(\tau) = - \langle
\hat{T}_\tau d(\tau) d^\dagger(0) \rangle$, where $\hat{T}_\tau$ is
the imaginary time-ordering operator. The corresponding spectral
function $A(\omega)$ is then constructed using the maximum entropy
method.~\cite{Jarrell/Gubernatis:1996}

\section{Results and Discussion}
\label{sec:results}

\subsection{SrVO$_3$}
\label{subsec:SVO}

We start by discussing the case of SrVO$_3$, where the $V^{4+}$ cation
exhibits a formal $d^1$ electron configuration. SrVO$_3$ is a rare
example of a material that exhibits an ideal cubic perovskite crystal
structure, i.e., without any symmetry-lowering distortions. SrVO$_3$
is often regarded as a prototypical example for a ``correlated
metal'', i.e., a metallic system where the electron-electron repulsion
leads to pronounced mass enhancement and narrowing of the
quasiparticle bands. Due to its simple crystal structure and the fact
that the $t_{2g}$ bands are well isolated from other bands at higher
and lower energies it is often used to test new DFT+DMFT
implementations and their
extensions.~\cite{Lechermann_et_al:2006,Karolak_et_al:2011}

\begin{figure}
\includegraphics[width=1.00\columnwidth]{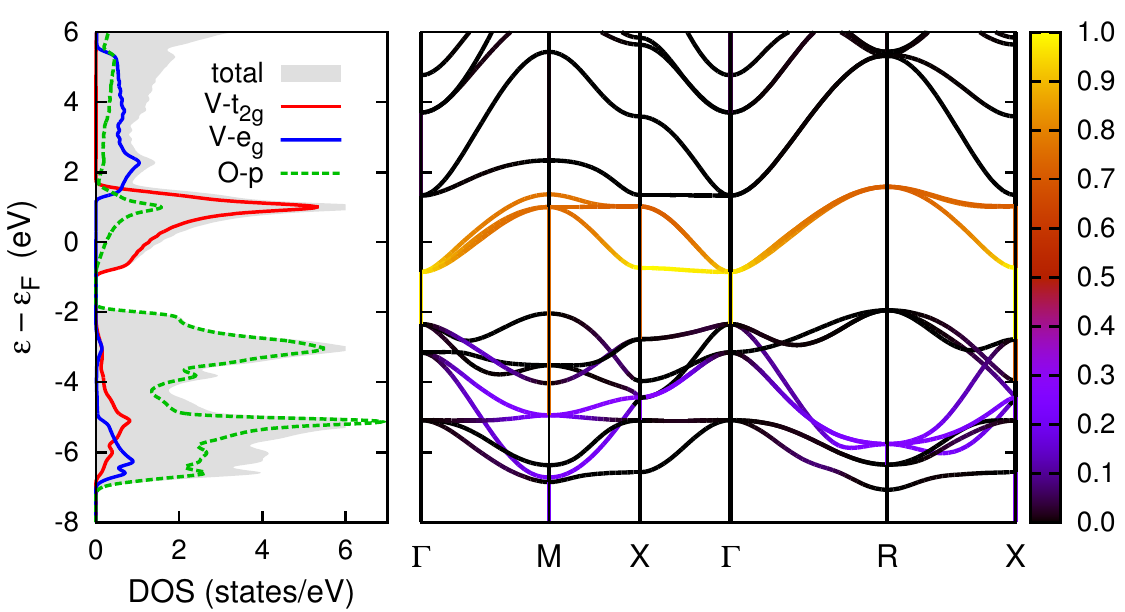}
\caption{Electronic structure of unstrained SrVO$_3$ around the Fermi
  level, $\varepsilon_{\rm F}$, obtained from DFT.  Left panel: Total
  DOS (gray filled area) and partial DOSs (lines) projected onto
  atomic orbitals with V-$t_{2g}$ (solid red), V-$e_{g}$ (solid blue),
  and O-$p$ character (dashed green).  Right panel: Kohn-Sham
  eigenvalues along high-symmetry lines in $k$-space.  The color
  represents the amount of V-$t_{2g}$ character within each bands
  (color scale on the right), with a value of 1 (0) corresponding to
  maximal (zero) overlap of the corresponding Bloch wave-function with
  the V-$t_{2g}$ atomic orbitals.}
\label{fig:SVO-bands}
\end{figure}

The total and projected densities of states (DOS) close to the Fermi
energy, $\varepsilon_{\rm F}$, for unstrained cubic SrVO$_3$, together
with its $k$-resolved band structure, are shown in
Fig.~\ref{fig:SVO-bands}. It can be seen that there are indeed three
partially-filled bands with strong atomic V-$t_{2g}$ character that
are clearly separated from bands at lower (higher) energies with
dominant O-$p$ (V-$e_g$) character.

We now relax the out-of-plane lattice constant of SrVO$_3$ for fixed
in-plane lattice parameters, which are varied by $\pm 4$\,\% around
the obtained cubic equilibrium lattice constant ($a_0 =
3.855$\,\AA). Due to the high symmetry
of SrVO$_3$, with no octahedral rotations, there are no free internal
structural parameters. For each strained structure, we then construct
MLWFs corresponding to the three V-$t_{2g}$ bands, starting from
initial projections on atomic $t_{2g}$ orbitals centered at the V
sites. The resulting Wannier orbitals closely resemble the ones of
Ref.~\onlinecite{Lechermann_et_al:2006}, with strong $t_{2g}$
character on the central V atom and $p$-like ``tails'' located on the
surrounding oxygen ligands.

\begin{figure}
\includegraphics[width=1.0\columnwidth]{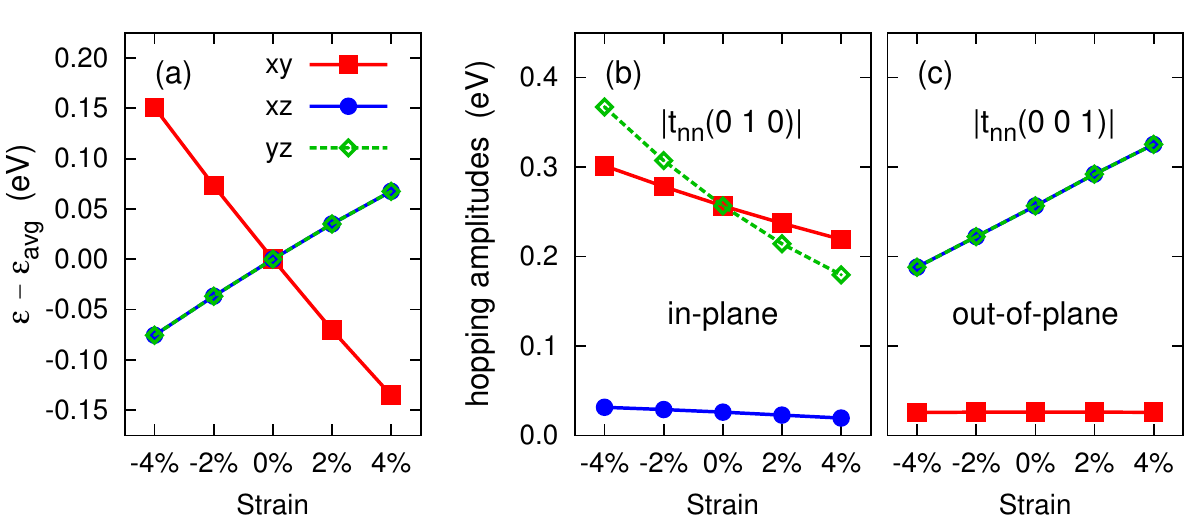}
\caption{Crystal-field splitting (a) and nearest neighbor hopping
  amplitudes along one of the two equivalent in-plane directions (b)
  and along the out-of-plane direction (c) in SrVO$_3$ as a function
  of epitaxial strain. The crystal-field levels in (a) are shown
  relative to the average $t_{2g}$ orbital energy for each strain.}
\label{fig:SVO-Hr}
\end{figure}

Fig.~\ref{fig:SVO-Hr} shows the strain-dependent crystal-field
splitting and nearest neighbor hopping amplitudes, obtained as on-site
and inter-site matrix elements, respectively, of the Kohn-Sham
Hamiltonians of the differently strained structures expressed in the
MLWF basis.  It can be seen that the crystal-field splitting indeed
follows the schematic picture sketched in
Fig.~\ref{fig:crystal-field-schematic}, with a splitting between
$d_{xz}$/$d_{yz}$ and $d_{xy}$ orbitals that is approximately linear
in the strain.
The calculated hopping parameters also follow the trends discussed in
Sec.~\ref{sec:general}, with the in-plane (out-of-plane) hopping
amplitudes decreasing (increasing) with strain. The strain dependence
of the in-plane $d_{xy}$ hopping is weaker than that of the
$d_{xz}$/$d_{yz}$ orbitals. The out-of-plane hopping for the $d_{xy}$
orbital and the in-plane hopping along $y$ for the $d_{xz}$ orbital
(and along $x$ for the $d_{yz}$ orbital) are very small, as expected
from the planar orientation of the $t_{2g}$ orbitals. Note that all
inter-orbital nearest neigbor hoppings are zero by symmetry.

Next, we perform DMFT calculations for the $t_{2g}$ bands using the
Hamiltonian expressed in MLWFs, where we add the electron-electron
interaction in the Slater-Kanamori form,
Eq.~\eqref{eq:slater-kanamori}. For each strained structure, we vary
the interaction parameter $U$, and identify the critical $U$ for the
Mott MIT by monitoring the value of the imaginary time Green's
function $G(\tau)$ at $\tau=\beta/2$. $G(\beta/2)$ is a measure of the
spectral density at the ``Fermi level'' $\omega=0$ (see, e.g.,
Ref.~\onlinecite{Fuchs_et_al:2011}):
\begin{equation}
A(\omega = 0) = - \frac{1}{\pi} \lim_{\beta \rightarrow \infty} {\rm
  Tr}[\beta G(\beta/2)] \quad .
\end{equation}

\begin{figure}
\includegraphics[width=1.0\columnwidth]{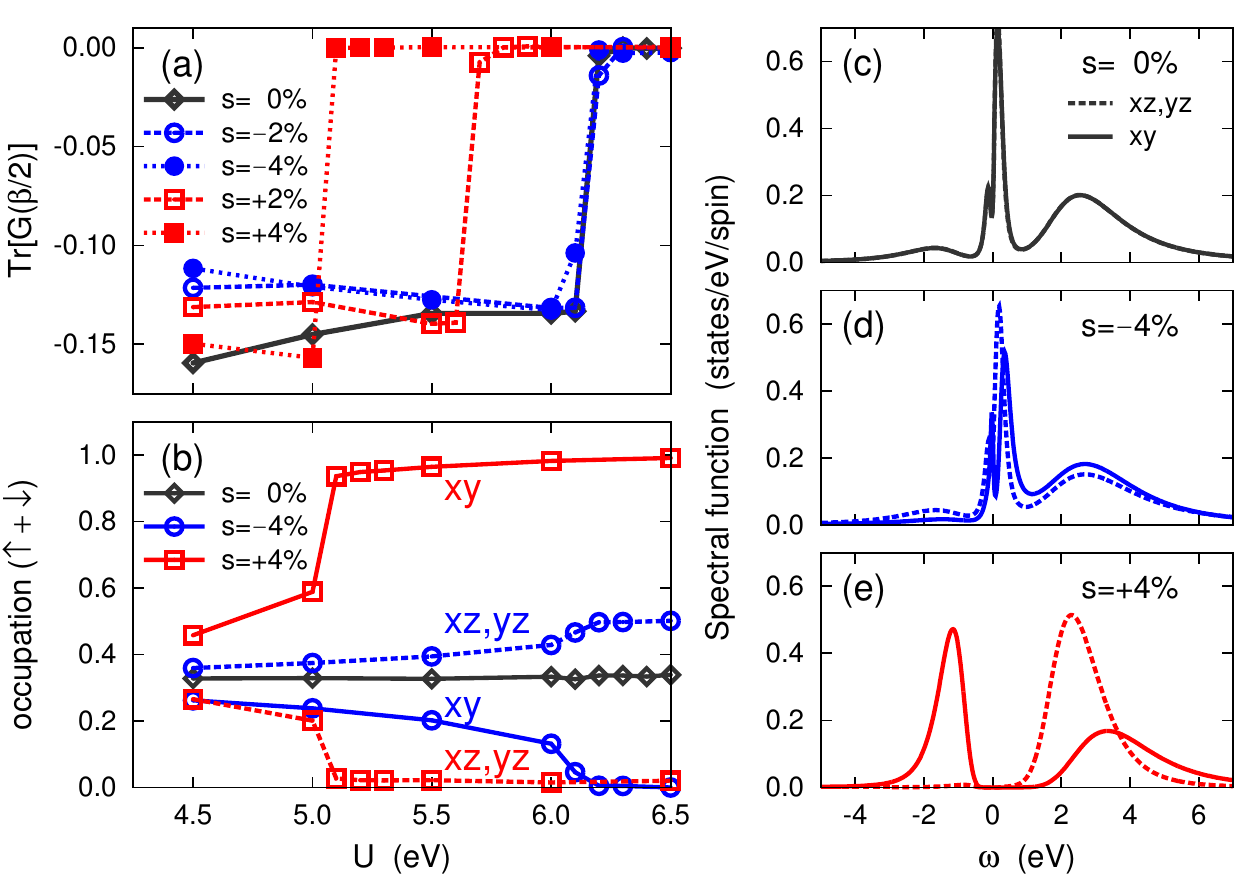}
\caption{DMFT results for unstrained SrVO$_3$ (black) and for SrVO$_3$
  under tensile (red) or compressive (blue) epitaxial strain.  (a)
  Trace of the Green’s function $G(\tau)$ at $\tau=\beta/2$ and (b)
  orbitally-resolved occupations, summed over both spin components,
  obtained from $-G(\beta)$, as a function of the interaction
  parameter $U$. The orbitally-resolved spectral function $A(\omega)$
  evaluated at $U=5.5$ eV is plotted as a function of $\omega$ for the
  unstrained case (c), under compressive strain (d), and under tensile
  strain (e). Different line styles in (b-e) differentiate between the
  $d_{xy}$ orbital (solid) and the degenerate $d_{xz}/d_{yz}$ orbitals
  (dashed).}
\label{fig:SVO-Gbetahalf}
\end{figure}

Fig.~\ref{fig:SVO-Gbetahalf}(a) shows the trace of $G(\beta/2)$ as
function of $U$ for different values of epitaxial strain. It can be
seen that in all cases $G(\beta/2)$ is nonzero for small $U$ (i.e.,
the system is metallic) and exhibits a transition to $G(\beta/2)
\approx 0$, i.e., to an insulating state, at some (strain-dependent)
critical value of $U = U_\text{MIT}$, which indicates the Mott
MIT. For unstrained SrVO$_3$ this transition occurs at $U_\text{MIT}
\approx 6$\,eV and this value is nearly unchanged under compressive
strain. In contrast, under tensile strain, there is a clear shift of
$U_\text{MIT}$ to lower values, with $U_\text{MIT} \approx 5$\,eV for
a tensile strain of 4\,\%.

The orbital occupations depicted in Fig.~\ref{fig:SVO-Gbetahalf}(b)
show that for zero strain all orbitals are equally populated both in
the metallic and in the insulating state (consistent with the cubic
symmetry of the system). For the strained systems, the crystal-field
splitting leads to an occupation imbalance between the energetically
higher- and lower-lying orbitals. While this occupation imbalance is
not very pronounced in the metallic state, the higher-lying orbital(s)
become completely empty in the insulating state.

It follows from the observed strain-induced shift in $U_\text{MIT}$,
that for a fixed value of $U$ in the range between $5\,\text{eV}
\lesssim U \lesssim 6\,\text{eV}$, a strain-induced metal-insulator
transition occurs at a $U$-dependent critical strain value $\leq
4$\,\%. Spectral functions for the case with $U=5.5$\,eV are shown in
Fig.~\ref{fig:SVO-Gbetahalf}(c)-(e). For a tensile strain of 4\,\% a
clear energy gap can be observed. On the other hand, under compressive
strain, the quasiparticle peak around $\omega=0$\,eV is slightly
broadened compared to the unstrained case, in particular for the
$d_{xy}$ orbital character. This is consistent with the expected
trends discussed in Sec.~\ref{subsec:hopping}

We note that a value of $U=5.5$\,eV is perhaps slightly too large for
SrVO$_3$, or at least it is at the upper end of the spectrum of $U$
values that are considered suitable to achieve a good description of
the electronic properties of this
material.~\cite{Pavarini_et_al:2004,Lechermann_et_al:2006,Dang_et_al:2014}
It is therefore unclear, whether large enough tensile strains can be
achieved in order to observe a strain-induced MIT in thin films of
SrVO$_3$. Furthermore, we note that it might be worthwhile to test,
whether the O-$p$ dominated bands starting at approximately 1\,eV
below the V-$t_{2g}$ bands (see Fig.~\ref{fig:SVO-bands}) also affect
the strain-dependence of the MIT, by including them into the DMFT
treatment of strained SrVO$_3$.
However, even if it might not be feasible to obtain insulating
SrVO$_3$ in epitaxial thin films under tensile strain, our results
clearly indicate that tensile epitaxial strain can lead to changes of
the quasiparticle effective mass and a partial suppression of orbital
fluctuations.

\begin{figure}
  \includegraphics[width=1.0\columnwidth]{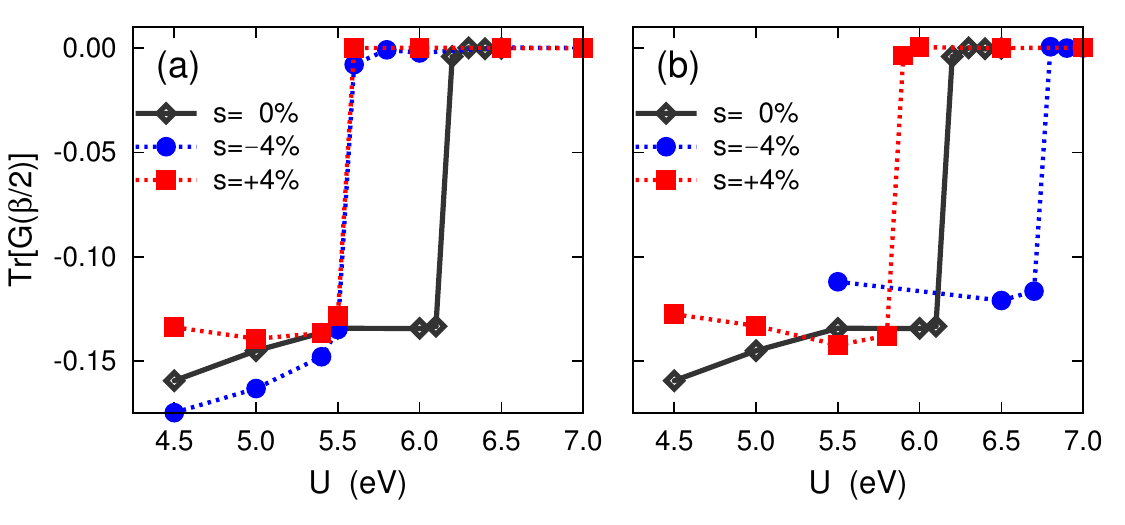}
  \caption{Separate effects of the strain-induced changes of the
    crystal-field splitting (a) and of the hopping amplitudes (b),
    i.e., in (a) the hopping amplitudes are fixed to their unstrained
    values and in (b) the same is done for the crystal-field
    splitting. The plots show the evolution of the trace of
    $G(\beta/2)$ as a function of $U$ under compressive strain (blue)
    and tensile strain (red) in comparison to the unstrained case
    (black).}
\label{fig:SVO-fake-strain}
\end{figure}

The strong decrease of $U_\text{MIT}$ under tensile strain is
consistent with the general trends expected both from the
strain-induced changes in the crystal-field splitting and in the
hopping amplitudes, whereas the absence of any shift under compressive
strain indicates a possible cancelation between the opposing trends
related to crystal-field and hopping.
To further verify this picture, and to better isolate the separate
effects of the strain-induced changes of the hopping amplitudes from
those of the crystal-field splitting, we perform additional
calculations where we fix one of the two types of parameters to the
corresponding unstrained values and only use the strain-dependent
values for the respectively other parameter. The results are depicted
in Fig.~\ref{fig:SVO-fake-strain}.

From the calculations with fixed ``unstrained'' hopping parameters
(Fig.~\ref{fig:SVO-fake-strain}(a)) it can be seen that the
strain-induced crystal-field splitting decreases $U_\text{MIT}$ for
both tensile and compressive strain.  Interestingly, the shift of
$U_\text{MIT}$ is identical for $\pm 4$\,\% strain within the limits
of accuracy of our calculations. This is consistent with the previous
model calculations of Ref.~\onlinecite{Werner/Gull/Millis:2009}, where
the filling dependence of the Mott MIT in the three-orbital model has
been studied, and the same critical chemical potential for the
destruction of the insulating state with one electron per site has
been found for both signs of the crystal-field splitting.

On the other hand, Fig.~\ref{fig:SVO-fake-strain}(b) reveals that the
strain-induced changes in the hopping parameters do affect
$U_\text{MIT}$ in different ways for compressive and tensile strain.
The observed trends are the ones expected from the discussion in
Sec.~\ref{subsec:hopping}, i.e., compressive (tensile) strain
increases (decreases) $U_\text{MIT}$ and thus favors the metallic
(insulating) phase.

These results confirm that for tensile strain the effects resulting
from the strain-induced changes in the crystal-field splitting and
hopping parameters cooperate, leading to a pronounced shift of
$U_\text{MIT}$ to lower values, i.e., the system moves closer to the
MIT. In contrast, under compressive strain the effect of the
crystal-field splitting counteracts the effect stemming from the
changes in the hopping amplitudes, leaving $U_\text{MIT}$ essentially
unaffected. This also shows that for a proper understanding of strain
effects on the MIT in correlated materials, the strain-induced changes
in both crystal-field and hopping parameters have to be taken into
account.

\subsection{LaTiO$_3$ and LaVO$_3$}

We now discuss the more complex cases of LaTiO$_3$ and LaVO$_3$. In
our previous work, we have already demonstrated that the Mott MIT in
the $d^1$ system LaTiO$_3$ is strongly affected by strain, with
tensile strain reinforcing the Mott-insulating character of LaTiO$_3$,
and a transition to the metallic state under compressive strain of
\mbox{1-2\,\%}.~\cite{Dymkowski/Ederer:2014} Thereby, the crucial
difference between the $d^1$ systems LaTiO$_3$ and SrVO$_3$ is the
presence of strong octahedral tilts in LaTiO$_3$, which lower the
space group symmetry to orthorhombic $Pbnm$ and distort the TM-O-TM
bond angles (see Fig.~\ref{fig:Pbnm}). This distortion of the ideal
cubic perovskite structure also leads to a pronounced crystal-field
splitting between the $t_{2g}$ orbitals \emph{already} for zero strain
and a more complex strain dependence of the hopping parameters
compared to SrVO$_3$.~\cite{Dymkowski/Ederer:2014}

In contrast, for the $d^2$ system LaVO$_3$, we found that the critical
$U$ for the Mott MIT is less affected by epitaxial
strain.~\cite{Sclauzero/Ederer:2015} Tensile strain leads to a
moderate decrease of $U_\text{MIT}$ in LaVO$_3$, i.e., strengthening
the insulating state, whereas compressive strain has nearly no effect
on $U_\text{MIT}$ (even though it has a noticeable effect on the
orbital polarization among the $t_{2g}$ orbitals). Thus, qualitatively
the trends in LaVO$_3$ are similar to the case of SrVO$_3$ discussed
in the preceding section. The very weak effect of compressive strain
on the MIT in LaVO$_3$ has been attributed to opposing effects of the
strain-induced changes in crystal-field splitting and
bandwidth,~\cite{Sclauzero/Ederer:2015} in analogy to the discussion
in Secs.~\ref{subsec:summary} and \ref{subsec:SVO}.

\begin{figure}
\includegraphics[width=1.0\columnwidth]{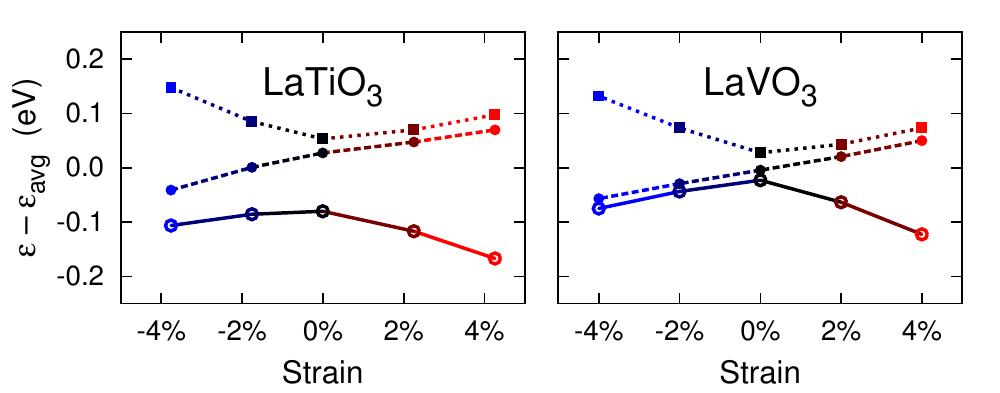}
\caption{Strain-dependent crystal-field levels of the three TM
  $t_{2g}$ orbitals for LaTiO$_3$ (left) and LaVO$_3$ (right).  For
  each strain, the orbital levels are shown with respect to the
  corresponding average value $\varepsilon_{\rm avg}$. The data for
  LaTiO$_3$ is reproduced from
  Ref.~\onlinecite{Dymkowski/Ederer:2014}.}
\label{fig:LVO-LTO-CF}
\end{figure}

In Fig.~\ref{fig:LVO-LTO-CF} we compare the calculated crystal-field
splitting between the three $t_{2g}$ orbitals of the TM cations in
LaVO$_3$ and LaTiO$_3$. The corresponding energies are obtained as
eigenvalues of the on-site part of the Kohn Sham Hamiltonian in the
basis of MLWFs, i.e., $\varepsilon_{nm}$ in Eq.~\eqref{eq:TB}. It can
be seen that in LaVO$_3$ the strain dependence of the crystal-field
splitting follows rather closely the schematic picture shown in
Fig.~\ref{fig:crystal-field-schematic}, and also observed for SrVO$_3$
in Fig.~\ref{fig:SVO-Hr}. Thus, in LaVO$_3$, the octahedral tilt
distortion results only in a weak splitting between the three $t_{2g}$
orbitals (see also
Ref.~\onlinecite{DeRaychaudhury/Pavarini/Andersen:2007}).

In contrast, the effect of the octahedral tilts is much stronger in
LaTiO$_3$, which exhibits a large crystal-field splitting already in
the unstrained state (see left side of
Fig.~\ref{fig:LVO-LTO-CF}).~\cite{Dymkowski/Ederer:2014,Pavarini_et_al:2004}
For zero strain, the splitting between the three $t_{2g}$ orbitals
resembles the tensile strain case in the schematic picture, with two
(nearly degenerate) orbitals at higher energies and one orbital at
lower energy. The corresponding splitting is further increased under
tensile strain. On the other hand, applying compressive strain reduces
the splitting between the lowest and second-lowest orbital and shifts
the highest-lying orbital further up in energy. This behavior was
already discussed in Ref.~\onlinecite{Dymkowski/Ederer:2014}.

\begin{figure}
  \includegraphics[width=1.0\columnwidth]{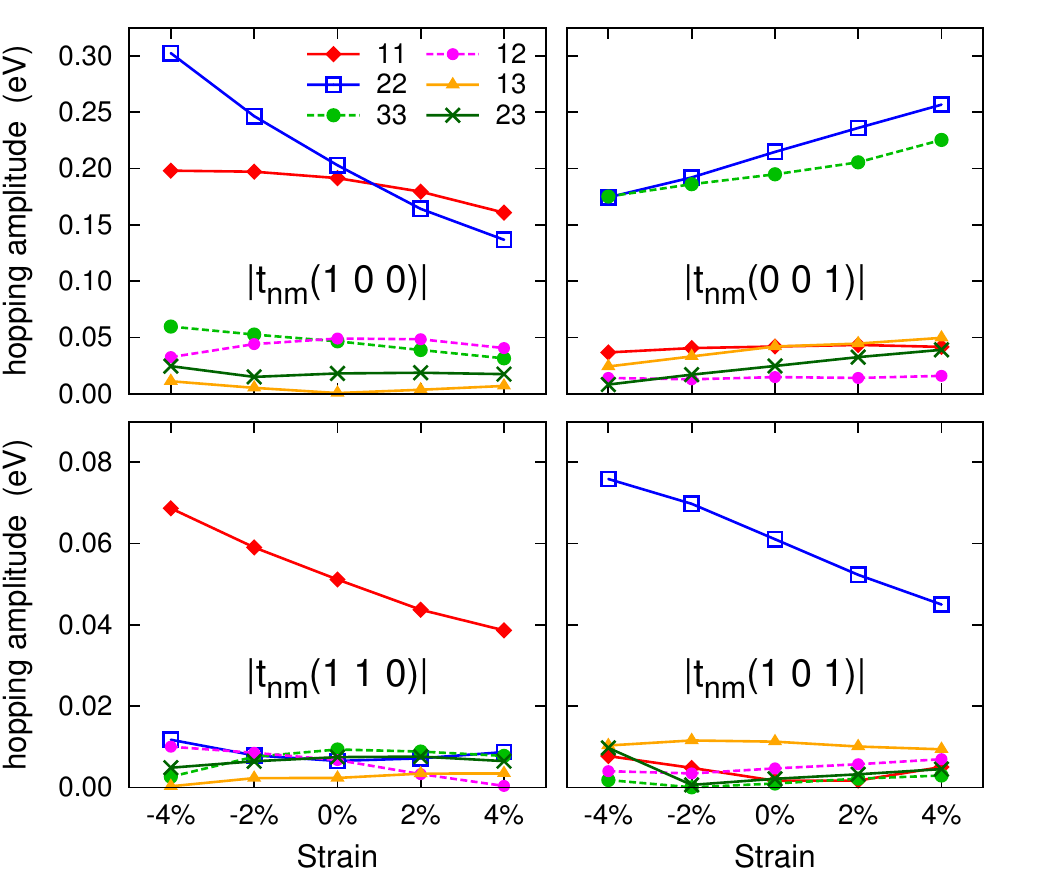}
\caption{Strain dependence of nearest neighbor (NN, top) and
  next-nearest neighbor (NNN, bottom) hopping amplitudes in
  LaVO$_3$. The left panels corresponds to in-plane hoppings (along
  $x$ for NN and along $x+y$ for NNN), the right panels correspond to
  out-of-plane hoppings ($z$ for NN and $x+z$ for NNN). Orbital ``1''
  roughly corresponds to $d_{xy}$ and is oriented mostly in-plane,
  while orbitals ``2'' and ``3'' derive from $d_{xz}$ and $d_{yz}$,
  respectively (see text).}
\label{fig:LVO-hoppings}
\end{figure}

Fig.~\ref{fig:LVO-hoppings} shows the calculated hopping amplitudes
for nearest neighbor (NN) and next-nearest neighbor (NNN) hopping in
LaVO$_3$ as function of epitaxial strain. It can be seen that the
dominant in-plane NN hoppings are decreasing (increasing) under
tensile (compressive strain) and vice versa for the out-of-plane
hoppings. This closely resembles the trends observed for SrVO$_3$ in
Fig.~\ref{fig:SVO-Hr}, and demonstrates that the changes in the
hopping amplitudes are indeed dominated by the changes in the TM-O
bond lengths and not by the changes in the TM-O-TM bond angles (see
discussion in Sec.~\ref{sec:general}).
The MLWFs in LaVO$_3$ reflect approximately the shape and orientation
of the $t_{2g}$ orbitals used to construct their initial projections
($d_{xy}$, $d_{xz}$, and $d_{yz}$ for orbitals 1, 2, and 3,
respectively), even though this is not enforced by symmetry. Thus, the
off-diagonal NN hoppings as well as, e.g., the hopping along $x$ for
the $d_{yz}$-derived MLWF remain small.
The dominant NNN hopping corresponds to the diagonal directions within
the plane in which the corresponding MLWFs are oriented, and are about
a factor 4 smaller than the dominant NN hoppings. They exhibit a
similar strain dependence as the dominant NN in-plane hoppings.

Thus, as for the crystal-field splitting, the effect of the octahedral
tilts on the hopping amplitudes is relatively weak in LaVO$_3$. The
main effect is simply an overall reduction of the dominant hopping
amplitudes. Calculations for LaVO$_3$ in a hypothetical ideal cubic
pervskite structure (not shown here) lead to hopping amplitudes that
are about 25\,\% larger than for $Pbnm$ LaVO$_3$, and are comparable
to the hopping amplitudes obtained for SrVO$_3$.
We also note that such a hypothetical cubic LaVO$_3$ has a critial $U$
for the Mott MIT of about 5.1~eV, i.e., almost 1~eV larger than in
$Pbnm$-LaVO$_3$.~\cite{Sclauzero/Ederer:2015} Thus, cubic LaVO$_3$
would be very close to the MIT or even be metallic. Therefore, it
appears that, similar to LaTiO$_3$,~\cite{Pavarini_et_al:2004} the
insulating state in LaVO$_3$ is stabilized by the octahedral tilt
distortion. However, in contrast to LaTiO$_3$, the crucial effect in
LaVO$_3$ is not a suppression of orbital fluctuations due to a strong
crystal-field splitting, but rather the resulting reduction of hopping
amplitudes.

\begin{figure}
\includegraphics[width=1.0\columnwidth]{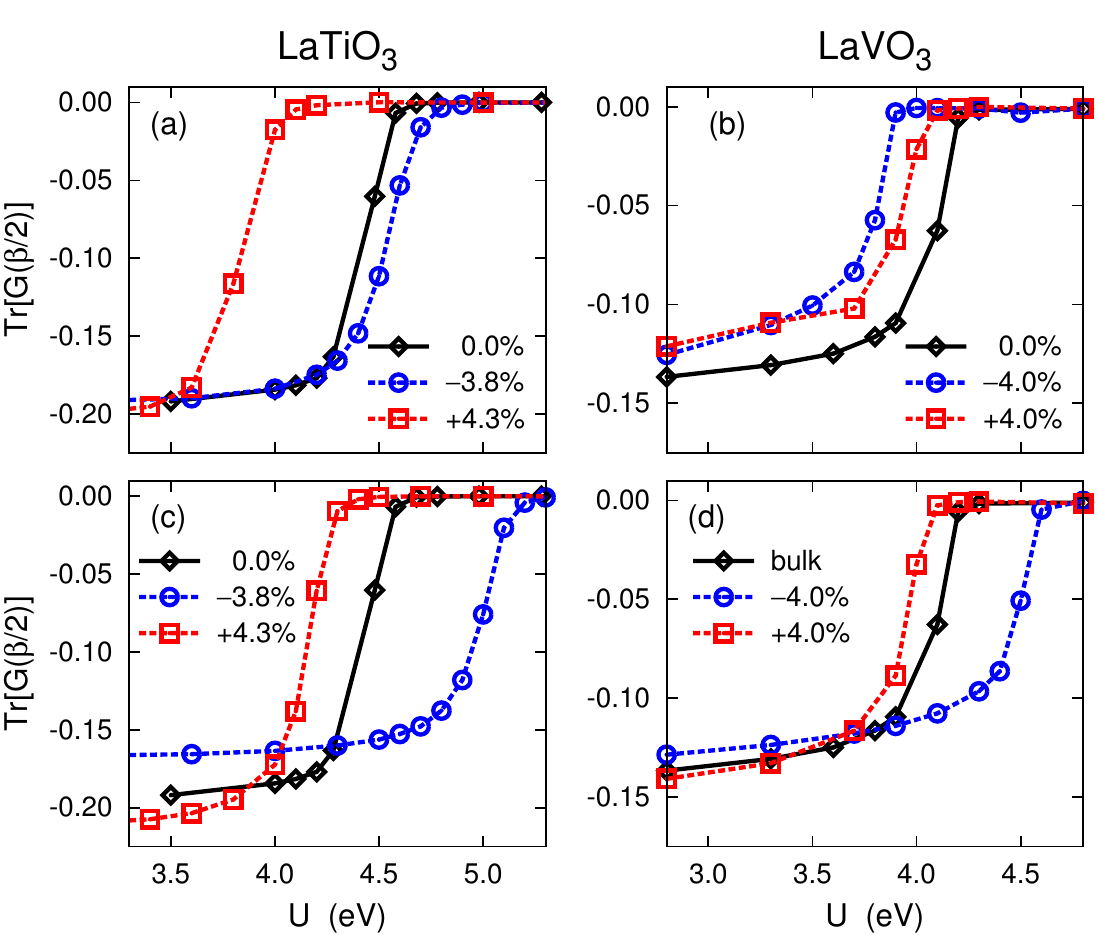}
\caption{Separate effects of the strain-dependent crystal-field
  splitting, i.e., hopping parameters fixed to the unstrained values,
  (top) and of the strain-dependent hopping parameters, i.e., crystal
  field fixed to the unstrained values, (bottom) for LaTiO$_3$ (left
  panels) and LaVO$_3$ (right panels).}
\label{fig:LVO-LTO-Gbhalf}
\end{figure}

Next, we present further analysis of the effect of epitaxial strain on
the electronic properties of LaTiO$_3$ and LaVO$_3$. In order to
clearly distinguish the effects resulting from the strain dependence
of the crystal-field splitting and from the strain dependence of the
hopping parameters, we perform DMFT calculations for strained
LaTiO$_3$ and LaVO$_3$ in the same way as described in
Sec.~\ref{subsec:SVO} for SrVO$_3$, i.e., we fix either the
crystal-field or the hopping parameters to their unstrained values
while using the strain-dependent values for the respectively other
type of parameter. The corresponding results, depicting the evolution
of ${\rm Tr}\,G(\beta/2)$ as function of $U$ for each case, are
summarized in Fig.~\ref{fig:LVO-LTO-Gbhalf}.

It can be seen that the effect of the strain-dependent hopping
parameters on $U_\text{MIT}$ (bottom panels in
Fig.~\ref{fig:LVO-LTO-Gbhalf}) is similar in both LaTiO$_3$ and
LaVO$_3$, and follows the general trends discussed in
Sec.~\ref{subsec:hopping} and also observed for SrVO$_3$ in
Fig.~\ref{fig:SVO-fake-strain}. Tensile (compressive) strain shifts
$U_\text{MIT}$ to lower (higher) values.  These shifts of
$U_\text{MIT}$ are somewhat larger in LaTiO$_3$ than in LaVO$_3$.

In contrast, the effect of the strain-induced crystal-field splitting
(top panels in Fig.~\ref{fig:LVO-LTO-Gbhalf}) is qualitatively
different in LaVO$_3$ and LaTiO$_3$.
In LaVO$_3$, $U_\text{MIT}$ is reduced under both compressive and
tensile strain, as expected from the strain-induced crystal-field
splitting (see Sec.~\ref{subsec:CF}) and similar to the case of
SrVO$_3$ (see Fig.~\ref{fig:SVO-Gbetahalf}). The reduction of
$U_\text{MIT}$ is stronger under compressive strain, which for a $d^2$
system leads to a situation with effective half-filling. However, for
both types of strain the decrease of $U_\text{MIT}$ in LaVO$_3$ is
significantly weaker than for SrVO$_3$.  

The case of LaTiO$_3$ is rather different. The critical $U$ for the
MIT is strongly reduced under tensile strain and is slightly increased
under compressive strain. This is due to the strong crystal-field
splitting already present in the unstrained state. Applying
compressive strain reduces the already existing crystal-field
splitting between the two energetically lowest orbitals and thus moves
the system closer to a situation with one electron in two degenerate
orbitals. This disfavors the insulating state and increases
$U_\text{MIT}$.~\cite{Gunnarsson/Koch/Martin:1996,Florens_et_al:2002}
On the other hand tensile strain strongly increases the splitting
between the lower-lying $d_{xy}$-like and the other two $t_{2g}$
orbitals, which has the ``normal'' effect of lowering $U_\text{MIT}$.

Thus, in LaTiO$_3$ the effects of crystal-field splitting and hoppings
cooperate for both types of strain. It is interesting to note, though,
that the increase of $U_\text{MIT}$ under compressive strain that is
caused by the strain-dependent hopping amplitudes is much larger than
the increase caused by the strain-dependent crystal-field splitting.
Therefore, the strain-induced changes of the hopping amplitudes appear
to be crucial for the insulator-to-metal transition induced in
LaTiO$_3$ under compressive strain.
In contrast, in LaVO$_3$ the effects of crystal-field splitting and
hopping amplitudes cooperate only for the case of tensile strain,
whereas they essentially cancel each other under compressive
strain. This is similar to the case of SrVO$_3$ discussed in
Sec.~\ref{subsec:SVO}, and confirms the interpretation/discussion
already given in Ref.~\onlinecite{Sclauzero/Ederer:2015}.

\section{Summary and Conclusions}

In summary, we have studied the effect of epitaxial strain on the Mott
MIT in various perovskite-structure TM oxides with $d^1$ and $d^2$
electron configuration of the TM cation, using first principles-based
DFT+DMFT calcuations. The MIT in these materials is governed by the
partially filled ``$t_{2g}$ bands'', derived mainly from the
corresponding orbitals of the TM cations.
We have analyzed the effect of epitaxial strain on these bands in
terms of strain-induced changes of the crystal-field splitting and
hopping amplitudes.

Our results show that the strain-induced changes in the hopping
amplitudes are qualitatively similar in all investigated systems. They
favor the metallic state under compressive strain and the insulating
state under tensile strain.
In contrast, the strain-induced crystal-field splitting will generally
always favor the insulating state, due to the reduced degeneracy of
the lowest energy orbitals. Consequently, the effects of crystal-field
and hopping amplitides usually cooperate under tensile strain,
favoring the insulating state, while they can effectively cancel each
other under compressive strain (see, e.g., the cases of SrVO$_3$ and
LaVO$_3$).

Strong octahedral tilts can modify these general trends, as in the
case of LaTiO$_3$. Here, the octahedral tilt distortion of the
perovskite structure leads to a strong crystal-field splitting already
in the unstrained state. This ``built-in'' crystal-field splitting
resembles the splitting that is otherwise induced under tensile
strain, and, as a result, compressive strain can induce a transition
to the metallic state by increasing the effective orbital degeneracy.
However, our calculations also demonstrate that the simultaneous
strain-induced increase of the in-plane hoppings leads to a much
stronger shift of the critical $U$ compared to the crystal-field
splitting alone.

Thus, our results show that both the strain-induced changes in the
crystal-field splitting and in the hopping amplitudes need to be
considered to correctly describe the effect of epitaxial
strain. Generally, the strain-induced changes in the hopping
amplitudes can be equally important as the strain-induced changes in
the crystal-field splitting for the overall behavior of the strained
material.

A particularly strong effect can be expected under tensile strain,
where the insulating state is strongly favored both by the
strain-induced changes in the crystal-field and hopping
parameters. This provides the possibility to move correlated metallic
systems such as SrVO$_3$ or CaVO$_3$ closer to the insulating state,
or even across the MIT, which is particularly interesting, e.g., for
the recently proposed applications of these correlated metals as
efficient transparent conductors.~\cite{Zhang_et_al:2015}

The impact of epitaxial strain should also be accounted for in view of
recent reports of MITs in ultra-thin films of
SrVO$_3$~\cite{Yoshimatsu_et_al:2010} and
CaVO$_3$.~\cite{Gu_et_al:2013} The insulating character observed in
these ultra-thin films has originally been attributed to a reduction
in bandwidth due to the dimensional crossover from a three-dimensional
metal to a two-dimensional
insulator.\cite{Yoshimatsu_et_al:2010,Gu_et_al:2013} In contrast,
Zhong {\it et al.} have recently argued that the crucial factor is
instead the crystal-field splitting between the $t_{2g}$ states that
is caused by the reduced symmetry in such ultra-thin films (which is
of the same type as the one induced under tensile epitaxial
strain).~\cite{Zhong_et_al:2015} Here, we show that the strain-induced
change in bandwidth is another important ingredient that can cooperate
or --- under compressive strain --- compete with the crystal-field
effect.

Finally, our work demonstrates that epitaxial strain indeed provides
an effective route to tune the strength of electronic correlations and
the vicinity to the Mott MIT in early TM perovskites. While further
systematic investigations are required to study the interplay between
strain and other effects occurring in thin films such as, e.g.,
interface effects, confinement, or defects, it is clear that strain is
a very important factor determining the properties of thin films and
heterostructures of correlated TM oxides.

\begin{acknowledgments}
This work was supported by ETH Zurich and the Swiss National Science
Foundation through grant No. 200021\_143265 and through the
NCCR-MARVEL. Calculations have been performed on the PASC cluster
``M\"onch'', hosted by the Swiss National Supercomputing Centre and
the ``Euler'' cluster of the ETH Zurich.
\end{acknowledgments}

\bibliography{references}

\end{document}